\documentclass[sigconf]{acmart}
\usepackage{subcaption}
\usepackage{comment}

\acmConference[EASE 2024]{The 28th International Conference on Evaluation and Assessment in Software Engineering}{18–21 June, 2024}{Salerno, Italy}
\AtBeginDocument{%
  \providecommand\BibTeX{{%
    \normalfont B\kern-0.5em{\scshape i\kern-0.25em b}\kern-0.8em\TeX}}}

\newcommand\fv[1]{\textcolor{orange}{#1}}
\newcommand\lm[1]{\textcolor{violet}{#1}}
\renewcommand\fv[1]{\textcolor{black}{#1}}
\renewcommand\lm[1]{\textcolor{black}{#1}}

\makeatletter
\def\@copyrightpermission{
  \hspace*{0mm}\includegraphics[width=2cm]{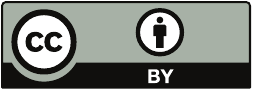}%
  \hspace*{2mm}\raisebox{2.5mm}[25pt][5pt]{%
          \parbox{\columnwidth}{\footnotesize This work is licensed under a Creative Commons \\ Attribution 4.0 International (CC BY 4.0) license.}%
  }\newline%
  The content in this pre-print is the same as in the CRC accepted for publication in:
}%
\makeatother
\acmDOI{10.1145/3661167.3661171}
\setcopyright{rightsretained}
\acmISBN{}
\copyrightyear{2024}
\acmYear{2024}
\acmConference[EASE 2024]{28th International Conference on Evaluation and Assessment in Software Engineering}{June 18--21, 2024}{Salerno, Italy}

\begin{document}

\title{Agent-Driven Automatic Software Improvement}

\author{Fernando Vallecillos Ruiz}
\affiliation{%
  \institution{Simula Research Laboratory}
  \city{Oslo}
  \country{Norway}}
\orcid{0000-0001-7213-3732}
\email{fernando@simula.no}
\renewcommand{\shortauthors}{Fernando Vallecillos Ruiz}

\begin{abstract}

With software maintenance accounting for 50\% of the cost of developing software, 
enhancing code quality and reliability has become more critical than ever.
In response to this challenge, this doctoral research proposal aims to explore innovative solutions by focusing on the deployment of agents powered by Large Language Models (LLMs) to perform software maintenance tasks.
The iterative nature of agents, which allows for continuous learning and adaptation, can help surpass common challenges in code generation.
One distinct challenge is the \emph{last-mile} problems, errors at the final stage of producing functionally and contextually relevant code.
Furthermore, this project aims to surpass the inherent limitations of current LLMs in source code through a collaborative framework where agents can correct and learn from each other's errors.
We aim to use the iterative feedback in these systems to further fine-tune the LLMs underlying the agents,
becoming better aligned to the task of automated software improvement.
Our main goal is to achieve a leap forward in the field of automatic software improvement by developing new tools and frameworks that can enhance the efficiency and reliability of software development.

\end{abstract}


\maketitle
\begin{acks}
Prof. Dr. Leon Moonen advises the author of this paper.
Prof. Dr. Leon Moonen is affiliated with Simula Research Laboratory and BI Norwegian Business School, Oslo, Norway -  email: \emph{leon@simula.no}.
This work is supported by the Research Council of Norway through the secureIT project (IKTPLUSS \#288787).
\end{acks}

\begin{figure}[t]
    \includegraphics[width=\columnwidth]{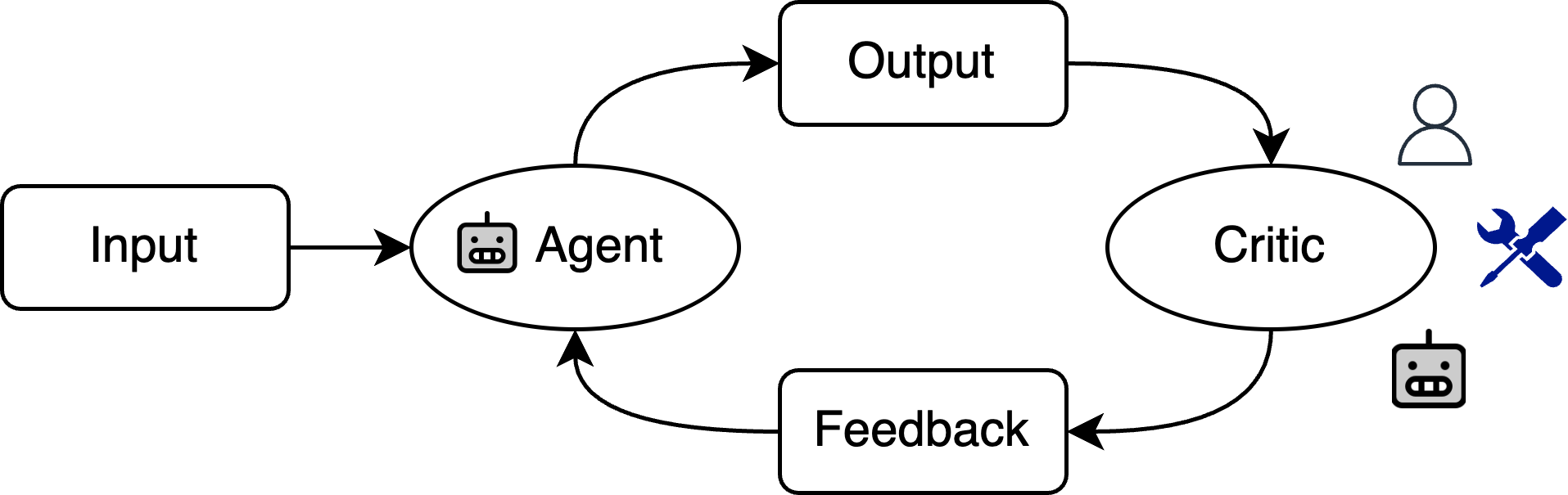}
    \caption{Conceptual framework for an LLM agent.}
    \label{fig:agent-framework}
    \Description{The conception framework of an LLM agent. The input is given to the agent. The agent generates an output which is received by a critic. The critic generates feedback, which is received by the agent. This completes the one cycle of iteration.}
\end{figure}
\section{Introduction}
In the current landscape of software development, 
the quantity of code is increasing at an unprecedented rate.
Although software facilitates progress across various sectors, it is plagued by issues such as bugs, vulnerabilities, and inefficiencies. 
These issues not only diminish performance and user experience, but also can pose substantial economic and security risks.
Remarkably, software debugging and maintenance consume half of the software development cycle~\cite{alaboudi2021:exploratory}.
The surge in Large Language Models (LLMs) has further aided this growth of code by providing developer tools to generate and auto-complete code.
LLMs, like human programmers, are prone to generating code with bugs and security vulnerabilities.
For example, Codex was shown to generate unsafe code 50\% of the time~\cite{pearce2021:asleep}.
Thus, the urgent need for robust 
code generation and maintenance
tools is underscored.

The paradigm shift towards agent-driven approaches is becoming increasingly evident~\cite{wang2023:survey}.
These 
approaches refer to intelligent agents that can perform tasks autonomously.
LLM-based agents, powered by an iterative use of an LLM, are designed to interact with the software environment, learn from said interactions, and make decisions autonomously.
The iterative use of LLMs has shown a significant performance improvement in complex debugging tasks~\cite{chen2023:teaching}.
The implications of enhancing code generation with iterative LLM-based agents are ambitious and clear. 
Through an iterative approach, which leverages continuous learning and adaptation abilities, agents are able to consistently outperform their one-shot counterparts~\cite{xia2023:keep, xia2023:conversational}.
Traditional programming skills may become less relevant, as the focus shifts towards areas such as system design and meta-programming.
Furthermore, a deeper understanding of the iterative nature of LLM-based agents could enhance their meta-learning capabilities, thereby expanding the scope and capabilities of LLMs even further.

The main objective of this project is to research the use of LLM-based agents to enhance the process of source code improvement.
Through state-of-the-art LLMs 
and in conjunction with collaborative agent frameworks, we aim to address primary aspects of source code quality: security, bug reduction, and efficiency optimization.
A simple version of the framework is shown in Figure \ref{fig:agent-framework} where the agent corresponds to an LLM, and the critic corresponds to an entity capable of providing feedback (such as a human, a tool or a different LLM).
The use of multiple LLMs, each serving as an expert in a sub-domain of software engineering, can help surpass existing benchmarks that have, to this day, remained unsolved by current technologies.
Furthermore, diverse fine-tuning and alignment techniques have been emerging in the previous years~\cite{zhang2023:instruction, shen2023:large}.
Our project, built on LLM-based agents, will iterate through cycles of code generation and revision, which runs parallel to some of these fine-tuning approaches.
These iterations can build the way for new fine-tuning techniques with the potential to significantly increase the quality of code synthesis.

\begin{figure*}
  \begin{subfigure}{0.15\textwidth}
    \includegraphics[width=\linewidth]{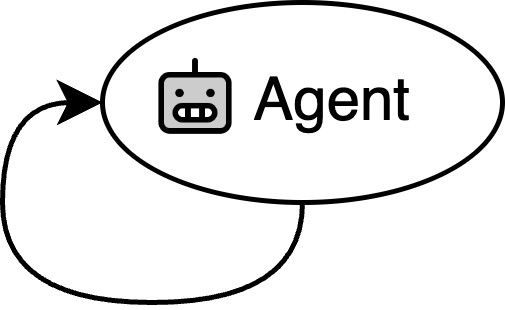}
    \caption{Single Agent}
    \label{fig:single-agent-interaction}
    \Description{One agent generates output and interacts only with itself.}
  \end{subfigure}%
  \quad \quad
  \begin{subfigure}{0.26\textwidth}
    \includegraphics[width=\linewidth]{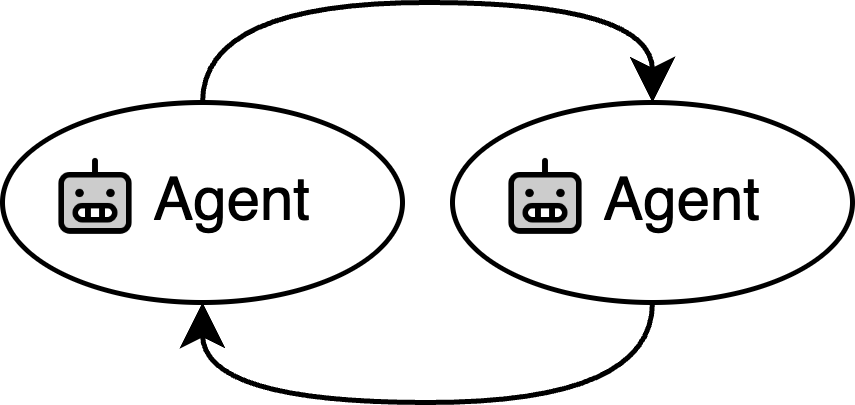}
    \caption{Multiple Agents}
    \label{fig:agent-agent-interaction}
    \Description{Two agents generate outputs and interact with the output of each other.}
  \end{subfigure}%
    \quad \quad
  \begin{subfigure}{0.26\textwidth}
    \includegraphics[width=\linewidth]{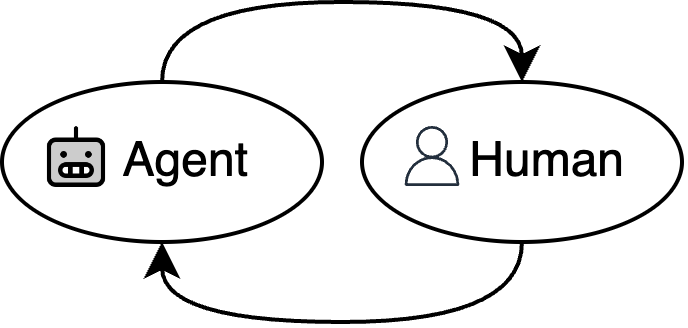}
    \caption{Human-in-the-Loop}
    \label{fig:agent-human-interaction}
    \Description{One agent and one human interact with the output of each other.}
  \end{subfigure}
  \caption{Scenarios of LLM-based agents interactions.}
    \label{fig:tye-agent-interaction}
    \Description{Three scenarios for the types of interaction of LLM-based agents.}
\end{figure*}

\section{Background and Challenges}
\subsection{Automatic Software Improvement}
The field of automatic software improvement has 
gained significant attention in recent years, driven by the growing need for efficient, error-free, and adaptive tools capable of handling complex software systems.
It encompasses a large number of techniques and methodologies aimed at reducing the amount of human intervention such as automatic program repair (APR)~\cite{zhang2023:survey, fan2023:automated}, security hardening~\cite{ji2018:coming, he2023:large}, or efficiency optimization~\cite{shypula2023:learning}.

Automatic software improvement tasks are commonly evaluated through public benchmarks.
The surge of software engineering benchmarks, driven by the threat of data leakage~\cite{wang2023:software, zheng2024:survey}, poses a problem towards standardization.
This growth leads to models being tested on just a fraction of the available benchmarks, increasing the difficulty of creating a comparison between models~\cite{zhou2023:don}.
Additionally, newer models will incorporate these new benchmarks into their training data, resulting in the creation of even more benchmarks to counteract the risk of data leakage.
This cycle presents a recurring challenge for model comparison and benchmark creation.

\subsection{Large Language Models for Code}

The capabilities of LLMs extend to a broad spectrum of tasks, including within the field of software engineering~\cite{minaee2024:large}, where their influence has been transformative.
Although they were originally developed for
Natural Language Processing (NLP), LLMs have transformed 
the field of 
software engineering by automating the generation and refinement of many of the processes~\cite{niu2023:empirical}.
For example, code generation, optimization, and debugging have been overturned by the use of LLM-based tools~\cite{zeng2022:extensive}.
In a short time, many developers have incorporated them into their daily routines.
These tools have significantly enhanced 
their productivity by automating a wide range of processes~\cite{kalliamvakou2022:research}.

Moreover, LLMs have their own set of challenges when applied to the field of software engineering.
Some recurrent issues are the generation of almost-correct code (\emph{last-mile errors})~\cite{bavishi2022:neurosymbolic}, the introduction of bugs~\cite{pearce2021:asleep} or even backdoors~\cite{hubinger2024:sleeper}.
These problems raise doubts about the reliability and security of the software generated using these tools.

The pursuit of improved outcomes, as well as the solution to the previous challenges, has resulted in many approaches.
There is a prevalent trend of ever-larger models, which usually correlates with better results~\cite{simon2023:more}.
This has been a temporal solution due to its non-sustainability~\cite{verdecchia2022:datacentric, hort2023:exploratory}.
However, a recent trend returns to smaller and more curated models that are capable of achieving similar or improved results compared to their larger counterparts~\cite{jiang2023:mistral, tunstall2023:zephyr, ma2024:era}
showing the potential to still refine our knowledge in LLMs.
Nonetheless, there is still a need to focus on more sustainable and diversified approaches to enhance their performance.

\subsection{Large Language Model-based Agents}
Agents, or entities that perform tasks autonomously, have become a focal point 
of
interest in many fields when combined with LLMs~\cite{wang2023:survey}.
It has been shown that LLMs benefit from task decomposition and multiple iterations, key parts in agent design strategies~\cite{radhakrishnan2023:question}.
A simplified version of the framework is shown in Figure \ref{fig:agent-framework}
where 
the agent (the LLM) obtains feedback from the critic to refine its response.
The combination of LLMs and agents creates tools with new capabilities and increased levels of intelligence.
The ability to decompose tasks, create plans, take decisions, and use tools makes agents a valuable system to tackle complex engineering tasks that were not able to be handled previously~\cite{xi2023:rise}.

Traditional one-shot methods, defined as approaches where a solution is attempted in a single-step, can be efficient in straightforward scenarios.
However, as the complexity of the code increases, these methods often fall short.
For example, last-mile bugs or complex error patterns may not be solved
through one-shot use of LLMs.
This tends to occur in tasks that require a deep level of contextual understanding and adaptability.

Iterative LLM-based agents offer a solution to these challenges by employing a multi-step repair process. 
Each iteration allows the agent to refine its understanding of the bug, consider additional contextual information, and adapt its solution accordingly. 
This approach not only increases the accuracy of the repairs, but also enhances 
its
ability to handle diverse and intricate bug patterns. 

Initial findings confirm the superiority of the iterative LLM-based approach compared to its one-shot counterpart~\cite{xia2023:keep, feldt2023:autonomous, xia2023:conversational}.
While this diversity is valuable, it also sheds light on the lack of direct, systematic comparison, and comprehensive analysis of these approaches.
Even without a clear understanding of these systems, multi-agents methodologies have already been employed for approaching the software engineering tasks successfully~\cite{hong2023:metagpt,qian2023:communicative, chen2023:gamegpt}. 

Agents may belong to a larger system designed to efficiently achieve one overall goal.
Following the approach proposed in previous work~\cite{xi2023:rise}, we discern three possible scenarios of interactions of an agent, as shown in Figure \ref{fig:tye-agent-interaction}.
\begin{enumerate}
    \item[(a)] Single agent: An LLM-based agent operates independently and interacts with tools or its own output.
    Agent improvement is achieved through continuous learning as a result of the self-feedback loop.
    \item[(b)] Multiple agents: Two or more LLM-based agents, each with distinct goals or objectives, operate collaboratively or competitively. 
    This scenario introduces the complexity of inter-agent communication and coordination, but it is particularly effective for complex problems requiring a multifaceted approach.
    For example, developing a game may require, among others, a project manager, a designer, and a developer~\cite{chen2023:gamegpt, qian2023:communicative}.
    \item[(c)] Human-in-the-loop: An LLM-based agent interacts with a human user, creating a hybrid approach that combines the computational power of the agent with the expert knowledge of the user.
    The human can provide domain-specific guidance to the LLM to align the outputs with the project's goals or constraints.
\end{enumerate}
This classification refers to individual interactions, rather than the overall structure of the system.
For example, an agent in a multi-agent system may engage in the three types of interactions shown in Figure \ref{fig:tye-agent-interaction}.
On the other hand, the system as a whole follows a structural architecture.
For example, it may be described as a hierarchical
or routing architecture.
These structural designs are crucial to determining how tasks and information are handled throughout the system.

\section{Expected Contributions}
This work can be divided into three phases that coincide with the main research questions.
We hypothesize that in the field of automated software improvement (a) LLM-based agents can perform better than one-shot LLM use; (b) multi-agent collaborative systems are able to consistently outperform single-agent systems; (c) the iterative communicative process can be used to fine-tune said LLMs.
The overall goal of the project is to design and implement agent-based frameworks for automatic software improvement.

Although previous work~\cite{talebirad2023:multiagent, guo2024:large, huang2024:understanding} has begun to address the potential of these agent-based systems, 
there remains a gap for comprehensive studies in various environments.
The design of LLM-based agents is closely related to their efficacy in code generation.
Therefore, it is crucial to investigate the factors that influence the accuracy and adaptability of these systems.
These factors cannot be understated given the increasing complexity of software systems and the increasing demand for automation in code generation.

The exploration of multi-agent code generation represents a significant challenge in software engineering research.
Our hypothesis that multi-agent systems outperform single-agent solutions in complex code-generating tasks is grounded on the need of multifaceted solutions from specialized experts.
The set of benchmarks CodexGLUE~\cite{lu2021:codexglue} exemplifies the wide range of code-generating tasks that require expert knowledge; therefore, a single model cannot excel at all of them.
A multi-agent approach would deploy agents designed or fine-tuned for a subset of tasks.
This approach leverages the strengths of individual models and allows for more adaptive and flexible systems capable of meeting the diverse demands of complex software projects.

From the architectural advantage of multi-agent systems, we posit that novel fine-tuning techniques shall be developed alongside.
Conventional fine-tuning approaches do not account for the dynamic and iterative process associated with code generation.
As such, we advocate for the development of fine-tuning methods that focus on the development of the agents' iterative learning and adaptation abilities.
These methods shall allow agents to efficiently refine their strategies in complex environments through continuous feedback loops.
The approach should lead to a system that iteratively converges towards optimal code generation.

This doctoral study aims to explore the nuances of LLM-based agent technologies and to develop novel methodologies for their fine-tuning, addressing the need for strategies that enhance their performance in code-generating related tasks.
Through the study of the dynamics between LLM-based agent frameworks, fine-tuning methodologies, and code generation tasks, this doctoral study will contribute towards the advance of effective strategies for automated code generation.
This research aims to generate insights able to provide a foundation for exploration and innovation in the application of LLMs to software engineering.

\section{Research questions and scientific challenges}
This PhD project will examine three main research questions:

\begin{itemize}
    \item[\textbf{RQ1.}] \textbf{Do tasks involving code improvement consistently benefit from the use of single-agent systems versus one-shot LLMs?}
\end{itemize}
    The emergence of recent studies using agents has suggested the superiority of iterative agents over one-shot LLMs for certain software engineering tasks such as APR or code synthesis~\cite{chen2023:teaching, xia2023:keep, xia2023:conversational}.
    While one-shot LLMs offer quick solutions, single-agent systems provide a more nuanced, iterative approach.
    Furthermore, factors such as task complexity or the integration of domain-specific knowledge may result in variations in the results~\cite{chen2023:teaching}.
    Therefore, it is essential to compare and create a framework of reference where we can identify the advantages of single-agent systems and for which specific tasks each method excels or falls short.

    The main challenge when confronting RQ1 is establishing an experimental setup with robust evaluation metrics that can fairly compare performance between single-agent systems and one-shot LLMs for code-generating tasks.
    This research should account for the differences between LLMs, their evolving nature, and the different application contexts.

    \begin{itemize}
    \item[\textbf{RQ2.}] \textbf{What are the synergistic effects of using multiple agents for code improvement tasks?}
    \end{itemize}
    The potential of multiple agents working simultaneously and interacting with each other remains an unexplored area~\cite{guo2024:large}.
    This research question investigates the effects of using multiple LLM-based agents in code improvement tasks, focusing on how their interaction may lead to enhanced performance compared to their single-agent counterpart.
    We will further concentrate on a different range of tasks, particularly in complex scenarios where the interaction of multiple expert-like systems could be essential to find an optimum solution.

    The main challenge associated with RQ2 is the complexity of orchestrating multi-agent systems and evaluating their performance, particularly with respect to their interactions and possible emergent behaviors.
    Understanding the influence of the agents chosen and their compatibility is essential to develop accurate evaluations of these systems.
    \begin{itemize}

    \item[\textbf{RQ3.}] \textbf{What novel fine-tuning techniques and methodologies can be created through the iterative process of code generation and revision %
    using LLM-based agents?}
        \end{itemize}
    New fine-tuning strategies have greatly impacted the use of LLMs.
    This research question seeks to develop a novel fine-tuning methodology that focuses on the iterative process of code generation and revision.
    \lm{The goal is to enhance the agent's capabilities even further by leveraging an iterative feedback loop for continuous improvement and adaptation to specific tasks.}
    This %
    can lead to,
    for example,
    more accurate and efficient generation by trying to shorten the number of iterations needed to achieve an optimum solution.

    The main challenge in addressing RQ3 lies in creating a dataset that integrates the iterative process of generation and revision, and developing and validating the new fine-tuning technique that can effectively incorporate it.
    The research not only involves multiple challenges in curating a dataset and developing a new methodology, but also accurately measuring the impact and effectiveness.

\section{Research Agenda}

\begin{figure}[t]
    \includegraphics[width=\columnwidth]{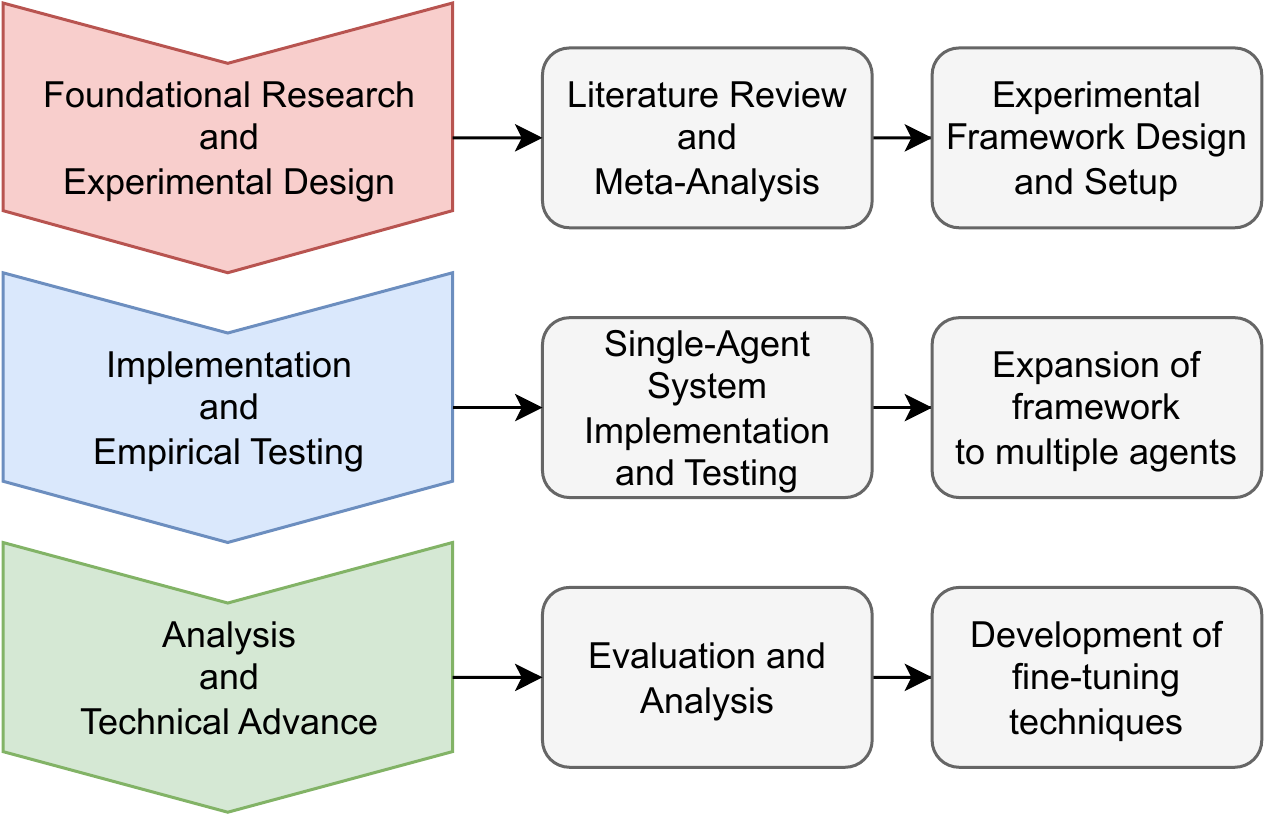}
    \caption{Overview of the research phases and the steps inside each phase.}
    \label{fig:agent-phases}
    \vspace*{-2ex}
    \Description{Overview of the research phases and steps in each phase. There are three main phases and two steps in each phase.}
\end{figure}

The work is divided into three phases, each containing two steps.
This division is depicted in Figure \ref{fig:agent-phases}.

\subsection{Phase 1: Foundational Research and Experimental Design}
\noindent\textbf{%
Step 1. Literature review and Meta Analysis:} 
The objective of this step is to systematically review research on LLMs and the technologies of LLM-based agents.
The rapid advancements and the growing scope of LLMs applications in the field of software engineering motivate the need for a structured exploration.

To achieve this first, we should gather studies from different sources, including journals, conference proceedings, and preprints, to capture the latest findings in the area.
This review should focus on code-generation related tasks and perform a meta-analysis to identify trends, gaps, opportunities, and metrics.
We will focus on studies related to code generation using LLMs, single-agent and multi-agent systems, and LLM-based agents.

\noindent\textbf{%
Step 2. Experiment Design and Setup:} 
This step's objective is to design experiments able to accurately measure the effects of different approaches on code-generating-related tasks.
The motivation behind this step is to ensure that the outcomes are robust, valid, and broadly applicable while providing insights into the efficacy of different approaches.

\fv{A crucial \lm{aspect} of the methodology will be to apply sensitivity analysis to understand how input variation affects the final results.
This emphasis aligns with our strategic decision not to compete directly with massive models but instead to optimize resources and focus on the nuances of performance under varying conditions.
}

To this end, we will develop protocols for experiments detailing the selected criteria for tasks, datasets, LLM configuration, and agents.
\fv{Specific metrics are yet to be determined, but they shall go beyond traditional metrics such as BLEU~\cite{papineni2002:bleu}.
These conventional metrics are still standard but have been shown to have clear limitations~\cite{mathur2020:tangled}.
New metrics such as BERTScore~\cite{zhang2020:bertscore} are being developed to improve the evaluation by focusing on semantic alignment rather than lexical similarity.
}
The chosen metrics are critical as they will serve as a baseline and will allow solid empirical investigation.
\fv{The protocols will be published in publicly accessible repositories to ensure transparency and allow for replication.}

\subsection{Phase 2: Implementation and Empirical Testing}
\noindent\textbf{%
Step 3. Implementation and Testing on Single Agents:}
Focused on single-agent systems, this step aims to implement and experiment across a variety of models and code-generating tasks, collecting information to evaluate their performance relative to one-shot LLMs.
Our driving force is the hypothesis that single-agent systems possess unique context-dependent properties.

\fv{%
The sensitivity analysis
can identify how variations in prompts or parameters influence the final results.
These variables have been shown to \lm{have great influence} in one-shot use of LLMs.
For example, modifying the prompt complexity and specificity, or the generation-related parameters such as temperature.
}

To achieve this, we will design and develop a single-agent system tailored for a specific code-generating task.
Then, we should conduct experiments in a controlled environment, i.e. in a deterministic manner, which will ensure reliability and applicability of the results.
The resulting data will be collected and analyzed to better understand the conditions under which single-agent systems outperform LLMs and when they should be applied.
This should result in a deeper understanding of LLM-based agents when applied to code-generating tasks.

\noindent\textbf{%
Step 4. Expansion of framework to Multiple Agents:}
The objective is to extend the previously implemented framework to handle multi-agent systems.
We believe that through collaborative efforts, multi-agent systems are capable of achieving better results than single-agent systems.

Towards this end, we will develop configurations for multi-agent systems focusing on role specialization and different interaction dynamics.
We should test the systems on complex code-generating tasks, measuring the impact of synergies and dynamics on the performance and efficiency of the tasks.
This new framework should conduct to a comprehensive study on the dynamics of the interactions and potential synergies and configuration in code-generating tasks.
These results will extend the previous comparison of the single-agent and one-shot systems, which should help identify key factors in the performance.

\subsection{Phase 3: Analysis and Methodological Advancements}
\noindent\textbf{%
Step 5. Evaluation and Analysis:}
The objective is to perform a comprehensive and thorough analysis of the experimental results.
In contrast to previous steps where we evaluate the agent framework through individual experiments and specific tasks, this step's goal is to produce a broader view.
The motivation is to synthesize the findings and draw conclusions on the efficacy of LLM-based agents in single and cooperative environments in code-generating-related tasks.%

To this end, we aggregate data from all experimental steps and applying statistical analysis can help uncover underlying patterns.
We extract patterns on the characteristics of the tasks and conditions that influence the success or failure of the approaches.
We then document the findings and reflect on the experimental design, lessons learned, and methodologies applied to note limitations and potential biases suggesting improvements for future research.
\fv{%
Other
work \lm{has} just shown improvement in performance.
\lm{However, we want to ensure that we perform a thorough analysis that can shed light on the effects of variables such as prompt, parameter, or randomness and their effect on results and applicability.}
}

\noindent\textbf{%
Step 6. Development of Fine-tuning Techniques:}
The final step focuses on leveraging the insights obtained from the previous steps to develop and test new fine-tuning techniques aimed at improving the performance of LLM-based agents.
This is motivated by the identified bottlenecks of the different systems suggested by the prior experiments.

This step will build upon the thorough analysis of the previous step to identify bottlenecks and areas of improvement.
A fine-tuning technique will be created adjusted to the target agent architecture.
The effectiveness of the techniques is tested through the evaluation metrics curated throughout the project to ensure its comparability and applicability across a diverse range of code-generating tasks and environments.

\subsection{Current Status}
We are currently engaged in Step 2 of the study.
The literature review has revealed a diverse range of studies on LLM-based agents, revealing a notable variability in research methodologies.
The goal of the reviewed literature is not to provide an in-depth analysis or to conduct subsequent experiments; therefore, we cannot determine which scenarios or conditions one approach outperforms the other.
While there is an emerging consensus on the benefits of single-agent systems compared to one-shot use of LLMs, the literature lacks a structured analysis on which scenarios are preferable or why they achieve better results.

In response, our intention in this step is to build a robust experimental framework able to handle one-shot use of LLMs, as well as LLM-based agents.
We start with APR as our initial code-generating task, and create a framework able to compare these two approaches.
The objective is to thoroughly analyze through different benchmarks in which cases APR may benefit from the use of agents.
This evaluation should extend beyond a simple benchmark comparison into an in-depth examination of the factors that result in single-agent systems improving the results of simple LLM usage.
Such an analysis can fill the gap on the underlying factors contributing to improved single-agent systems over traditional LLM implementations.

\fv{
\lm{We will initially focus on smaller and more manageable models.
This allows us to} systematically study fundamental optimizations within the models without the overhead associated with larger models.
By starting with smaller models, we 
are
able to conduct a more detailed analysis, which can provide insights into their bigger counterparts.
}

\section{Threats to Validity and Mitigation Strategies}
The threats to validity can be divided into four categories, structured cf. Wohlin et al.~\cite[Sec. 6.7 \& 6.8]{wohlin2000:experimentation}:

\textbf{Construct validity:} 
The metrics chosen can potentially positively bias the evaluation of the systems.
Therefore, we shall carefully choose a selection of widely-used metrics and benchmarks to approximate the performance to real-world scenarios.

\textbf{Internal validity:}
The iterative design and feedback process increase the importance of prompts, which can lead to vastly different results, complicating the attribution of outcomes to specific changes.
We shall conduct the experiments with variations of the prompts to incorporate this understanding into the design of the experiments and minimize its impact.

\textbf{External validity:}
The selection of code-generating tasks may not be representative to the broader range of possible tasks, therefore, potentially limiting the applicability of the findings.
We shall clearly state the environment used to evaluate the experiments to give a clear picture of the proven capabilities of the systems.

\textbf{Reliability:}
The complexity of the systems and the evaluation can lead to inconsistencies in the experimental setup and execution.
We shall provide replication packages with all the studies to ensure open science principles and make materials available to the scientific community.

\section{Conclusion}
This doctoral research proposal seeks to advance the domain of automatic software improvement through the capabilities of LLM-based agents.
We aim to solve common issues in one-shot usage of LLMs when generating code such as \emph{last-mile} problems.
We further overcome the limitations of current LLMs through a collaborative framework where agents can correct and learn from each other.
This framework leverages the agents' adaptability and learning skills to align with the increasing demands of automated software improvement.
Our objective is to design and develop innovative tools, frameworks, and techniques to improve the efficiency and reliability of software.
Substantial contributions towards reducing development cost, in addition to technical advancements, are to be expected from this proposal.
Ultimately, we expect this research to provide significant insights into the potential of LLM-based agent systems in the field of software development.

\bibliographystyle{ACM-Reference-Format}

\end{document}